\documentclass[journal]{IEEEtran}

\usepackage{url}
\makeatletter
\g@addto@macro{\UrlBreaks}{\UrlOrds}
\makeatother

\usepackage{mathtools}
\usepackage{tikz}

\usepackage[linesnumbered, ruled]{algorithm2e}

\usepackage[T1]{fontenc}
\usepackage{array}
\usepackage{booktabs}
\usepackage{latexsym}
\usepackage{tabularx}
\usepackage{enumitem}

\usepackage{color} 
\graphicspath{{./figures/}}

\newcommand{\ra}[1]{\renewcommand{\arraystretch}{#1}}

\newcommand{\card}[1]{\ensuremath{\left\|#1\right\|}}

\DeclareMathOperator*{\argmin}{arg\,min} % thin space, limits underneath in displays

\begin{document}
	\sloppy

\title{Towards Wi-Fi AP-Assisted Content Prefetching for On-Demand TV Series: A Reinforcement Learning Approach}


\author{Wen~Hu,
        Yichao~Jin,
		Yonggang~Wen,~\IEEEmembership{Senior~Member,~IEEE,}
		Zhi~Wang,~\IEEEmembership{Member,~IEEE,}
        and~Lifeng~Sun,~\IEEEmembership{Member,~IEEE}% <-this % stops a space
		}

\maketitle

\begin{abstract}
The emergence of smart Wi-Fi APs (Access Point), which are equipped with huge storage space, opens a new research area on how to utilize these resources at the edge network to improve users' quality of experience (QoE) (e.g., a short startup
delay and smooth playback). One important research interest in this area is content prefetching, which predicts and accurately fetches contents ahead of users' requests to shift the traffic away during peak periods. However, in practice, the different video watching patterns among users, and the varying network connection status lead to the time-varying server load, which eventually makes the content prefetching problem challenging. To understand this challenge, this paper first performs a large-scale measurement study on users' AP connection and TV series watching patterns using real-traces. Then, based on the obtained insights, we formulate the content prefetching problem as a Markov Decision Process (MDP). The objective is to strike a balance between the increased prefetching\&storage cost incurred by incorrect prediction and the reduced content download delay because of successful prediction. A learning-based approach is proposed to solve this problem and another three algorithms are adopted as baselines. In particular, first, we investigate the performance lower bound by using a random algorithm, and the upper bound by using an ideal offline approach. Then, we present a heuristic algorithm as another baseline. Finally, we design a reinforcement learning algorithm that is more practical to work in the online manner. Through extensive trace-based experiments, we demonstrate the performance gain of our design. Remarkably, our learning-based algorithm achieves a better precision and hit ratio (e.g., $80\%$) with about $70\%$ (resp. $50\%$) cost saving compared to the random (resp. heuristic) algorithm.
\end{abstract}


\section{introduction} \label{section:introduction}
\IEEEPARstart{T}{hese} years have witnessed the explosive growth of network traffic: Cisco~\cite{cisco} had predicted that Global Internet traffic in $2019$ will be equivalent to $64$ times the volume of the entire global Internet in 2005 and  consumer internet video traffic will be $80\%$ of all consumer Internet traffic in 2019, up from $64\%$ in 2014. Although existing video providers have adopted the Content Delivery Network (CDN) to help deliver videos to users across the world, its centralized way and the expensive deployment costs, however, make the conventional CDN not sufficient to provide satisfactory user-perceived QoE~\cite{khemmarat2012watching}. Besides, compared with traditional web objects, the size of video content is several orders of magnitude larger than that of web objects, making the storage space of centralized content servers exhaust much more quickly. This results that the temporal locality of videos can not be well exploited and users' QoE eventually degrades due to the lower video hit ratio. Moreover, the increasing availability of high quality videos further exacerbate the end users' QoE, e.g., high startup delay and frequent re-buffering events. 

To bridge the gap between the explosively increasing network traffic and the slow improvement on the performance of physical network infrastructures, a trend has emerged to shift the traffic at peak periods by prefetching them in advance. Some works~\cite{jiang2012orchestrating} have proposed to utilize the \emph{edge-devices} (e.g., set-top boxes, broadband gateway) to assist the video delivery. This approach becomes more promising thanks to the emergence of smart APs. Compared with tradition APs which only perform the data forwarding function, these smart APs (also called home router) are equipped with an OS and some storage devices (e.g., a hard disk drive or SD card). Furthermore, the enormous popularity of smart AP ($6.5$ M sales until 2015 in China~\cite{smart_ap}) enables these widely distributed resources to conduct the prefetching tasks.

% more video service providers (e.g., Youku, one of the largest video service provider in China) pays for its users to install smart routers

% Today, more and more smart APs  are deployed by either content provider sponsored (Youku pays for its users to install smart routers~\cite{youku_sponsor}) or user himself paid. .

However, in practice, the different video watching patterns among users, and the varying network connection status lead to the time-varying server load, which eventually makes pefetching efficiently challenging, i.e., we need to address the following challenges: i) What content should be prefetched to which AP ? ii) How many content items should be prefetched ? Intuitively, there is a trade-off between the prefetching costs and users' QoE, in the AP-assisted content prefetching problem. On one hand, aggressive prefetching strategy can guarantee a higher probability of hit ratio, thus a better users' QoE. On the other hand, more contents to prefetch, in turn, involves more monetary cost. Moreover, prefetching aggressively will incur the competition for the end user's downlink bandwidth with the current playback task, which will degrade the current viewing experience. 

To address these challenges, we perform a large-scale measurement study on users' AP connection and TV series watching patterns using real-traces. Based on the obtained insights, we formulate the content prefetching problem as a Markov Decision Process (MDP) and propose four algorithms to solve it. 

% Hence, it is crucial to schedule the content prefetching tasks with user behavior and server load awareness. 

% To address these challenges, we propose a users' behavior and server load aware sequential decision algorithm.

\begin{figure*}[t]
 		\begin{minipage}[t]{0.32\linewidth}
	 		\centering
				\includegraphics[width=\linewidth]{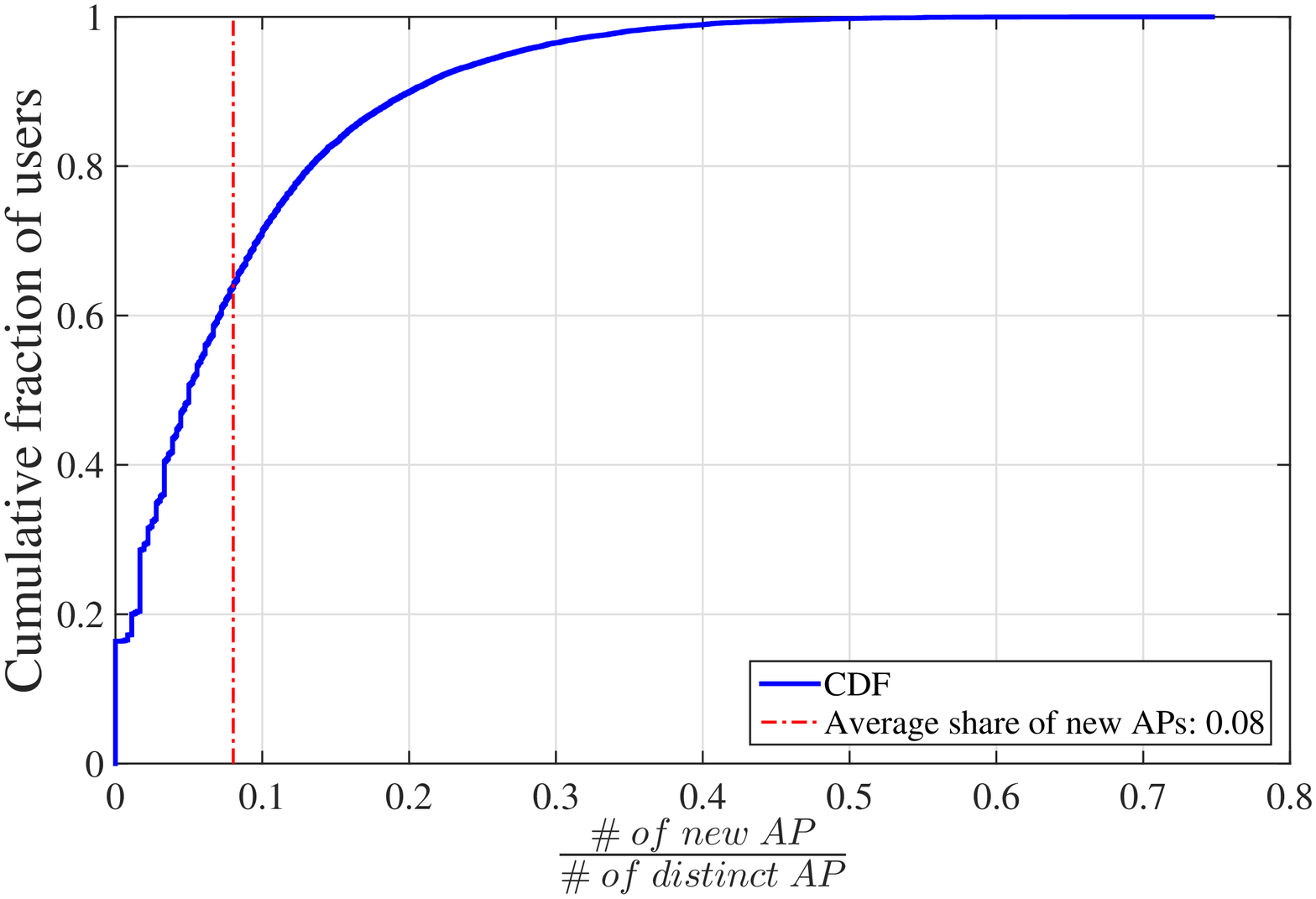}
	    			\caption{The fraction of the users vs. the ratio between new APs and distinct APs associated by each user per day.}
					\label{fig:new_ap}
 		\end{minipage}
 		\hfill
 		\begin{minipage}[t]{0.32\linewidth}
		    \centering
		        \includegraphics[width=\linewidth]{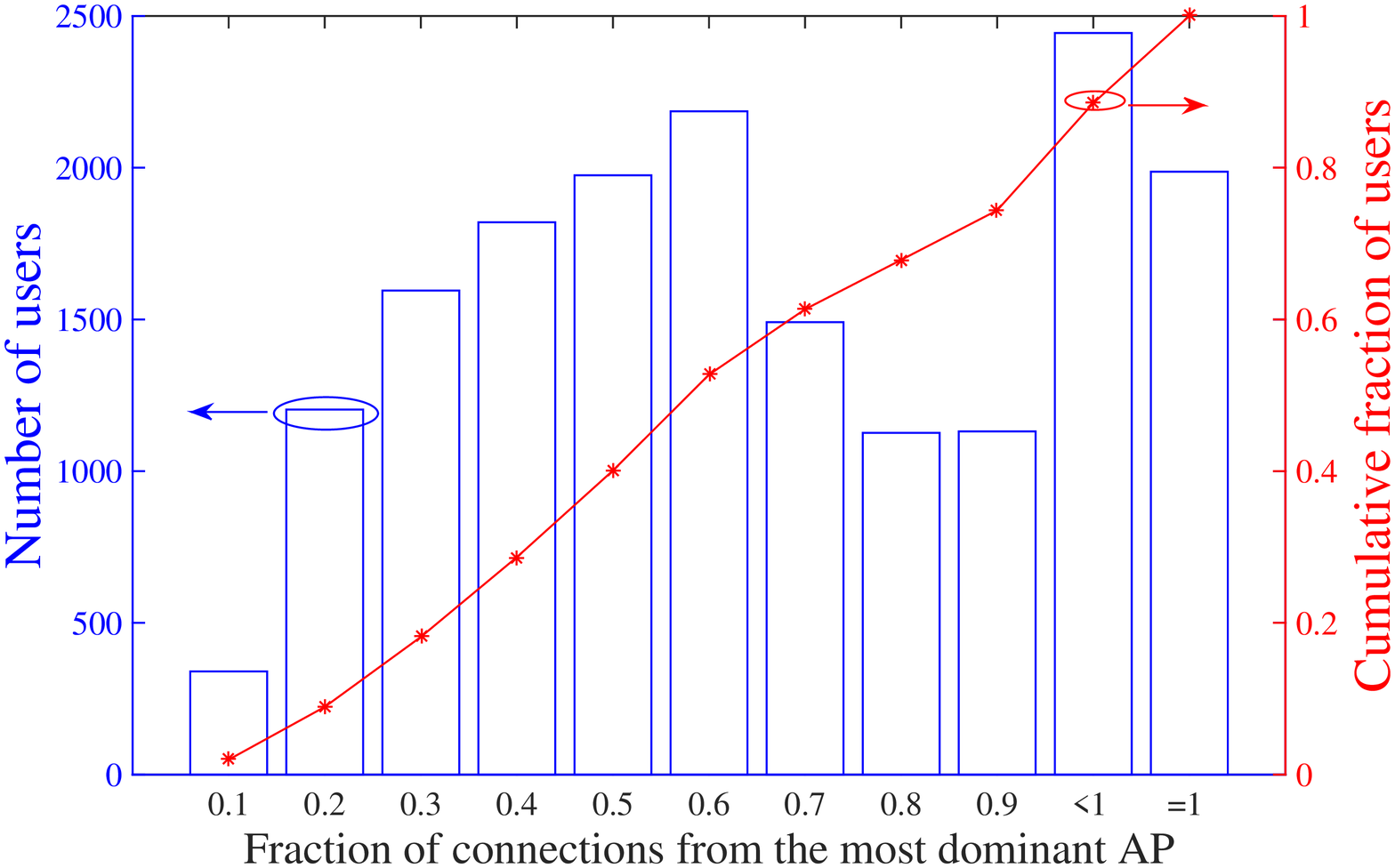}
		    \caption{The fraction of users vs. the contribution of each user's the most dominant AP.}
		     \label{fig:home_ap}
 		\end{minipage}
 		 \hfill
 		\begin{minipage}[t]{0.32\linewidth}
	    \centering
	        \includegraphics[width=\linewidth]{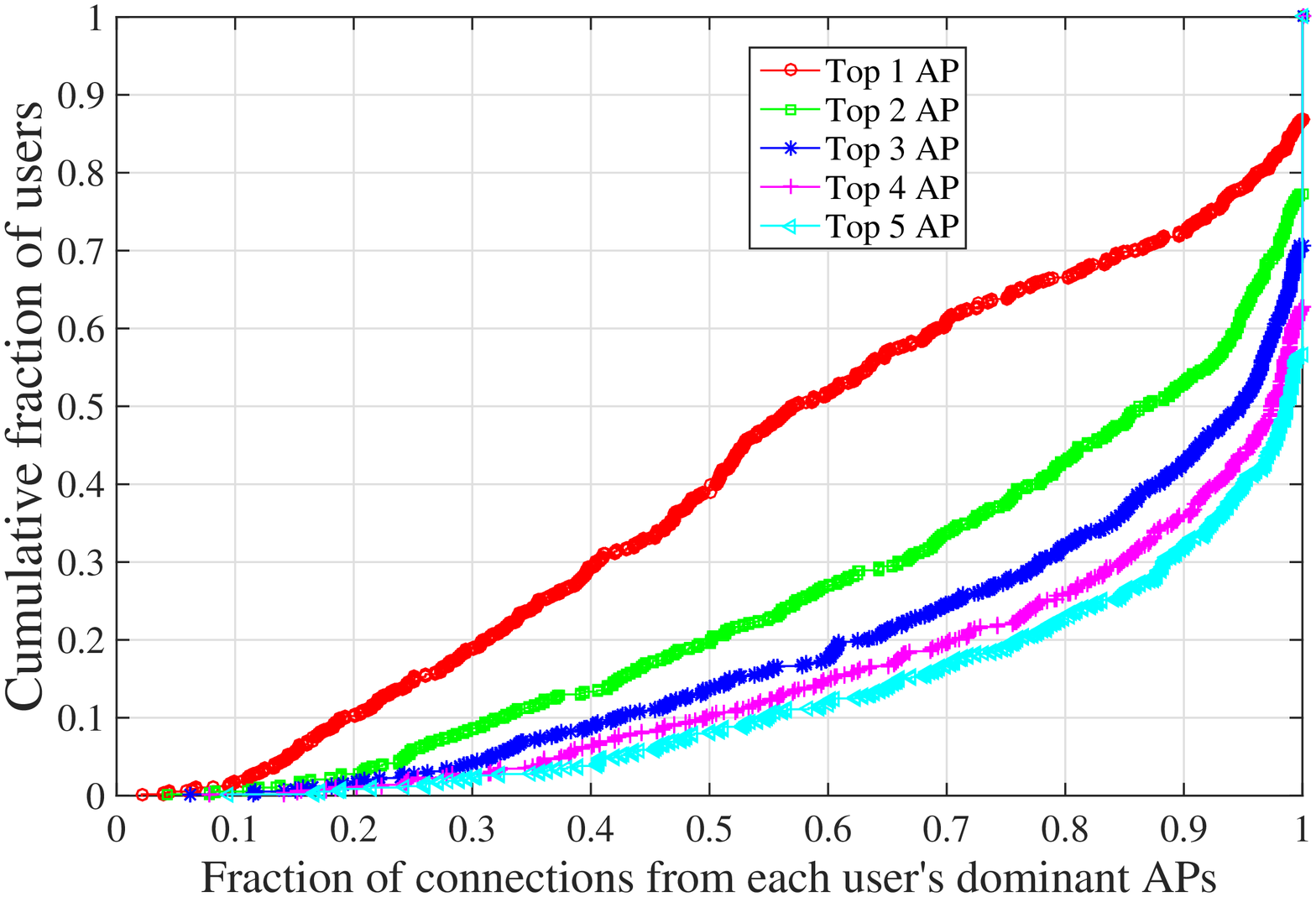}
	    \caption{The cumulative fraction of users vs. the contribution of each user's top-$K$ dominant APs.}
	     \label{fig:home_ap_topK}
 		\end{minipage}
 \end{figure*}
 
Our contributions in this paper are summarized as follows:

$\rhd$ We carry out large-scale measurement study on $270$ M user-AP association traces and $1.8$ M users' watching traces on $76$ K episodes from $8.5$ K TV series. The observations on users' AP connection and their TV series watching patterns are reported as follows: i) The user's connected APs are stable over time; ii) There are more than $60\%$ (\emph{resp.} $90\%$) users with the majority (e.g., over $50\%$) of their connections served by their top $1$ (\emph{resp.} $5$) AP(s); iii) Users tend to watch consecutive episodes in the same TV series. These observations motivate our work and provide valuable insights for our ensuing design.

$\rhd$ Based on our measurement studies, we propose the AP-assisted content prefetching paradigm and mathematically formulate the content prefetching problem as a Markov Decision Process. The objective is to balance the trade-off between various costs (including transmission, storage, increased latency and resource competition cost) and users' QoE.

$\rhd$ Under this framework, we propose four algorithms to schedule the content prefetching tasks in different periods. First, we obtain the lower and upper performance bound by proposing a random fixed and an offline algorithm, respectively. Then, we present a heuristic algorithm as another practical baseline. Finally, we propose a reinforcement learning algorithm to learn the history traces and schedule content prefetching accordingly in the online manner.

%  assuming users' watching behavior (i.e., the transition probability among episodes) is known a prior, we propose a semi-online algorithm. Finally

$\rhd$ Using trace-driven experiments, we further evaluate the performance of each algorithm. The results show that our design is adaptive to the server load and achieves about $70\%$ (resp. $50\%$) cost saving over the random (resp. heuristic) algorithm with high precision and hit ratios.

The rest of the paper is organized as follows. We present the measurement insights that motivate our design in Sec.~\ref{section:motivation}. We present the system architecture and the problem formulation in Sec.~\ref{section:formulation}. We propose our strategies in Sec.~\ref{section:algorithm}. We evaluate their performance in Sec.~\ref{section:evaluation}. We discuss the related works in Sec.~\ref{section:relatedwork}. Finally, we conclude this work in Sec.~\ref{section:conclusion}.

\section{Measurement and Analysis} \label{section:motivation}
In this section, we first introduce the datasets utilized in this paper. Then, we present some insights we learned from the exhaustive measurement study.

\subsection{Dataset}
\subsubsection{Traces of AP Connections}
We study users' AP association patterns using the dataset provided by Tencent\footnote{http://www.tencent.com. Hereinafter, The Wi-Fi Service Provider is referred to the service.}. The dataset contains $270$ M user-AP association traces during one month (March $2015$ - April $2015$). Each trace item records the user ID, the Basic Service Set Identifier (BSSID) of the AP, the timestamp of the association, and the location of the AP. Note that the user ID and the BSSID are unique for different users and APs, respectively. Thus we can identify each specific user and AP to learn the users' AP connection patterns.

\subsubsection{Traces of TV Series Sessions}
We investigate users' TV watching behavior, i.e., the transition among episodes in the same TV series, based on the traces provided by iQiyi\footnote{http://www.iqiyi.com. Hereinafter, The Video Service Provider is referred to the service.}, one of the most popular online video providers in China. The traces are collected from $1.8$ M users in a metropolitan city during $2$ weeks of May $2015$, containing $76$ K episodes on $8.5$ K TV series. In particular, the traces are recorded at the request level, i.e., the watching experience for an episode is recorded as one session. In each trace item, the following information is recorded: the user ID, the timestamp when the user starts to watch the video and the title of the episode. Based on these traces, we will study the users' TV watching behavior.

\subsection{User-AP Connection Pattern} 
 
First of all, we calculate the the ratio between the number of new APs, which have not been connected previously, and that of distinct APs associated by each user per day. We plot the cumulative distribution of aforementioned ratio in Fig.~\ref{fig:new_ap}. It is clearly that the number of new APs is much smaller than that of distinct APs. In particular, about $80\%$ users visit less than $20\%$ new APs per day and the average share of new APs is only $8\%$. 

Next, we explore the users' preference to a AP, which is defined as the ratio between the connections served by one AP and the total connections issued by a user, in Fig.~\ref{fig:home_ap}. In the analysis, we focus on the users who access network via APs at least once a day, and filter others. The bar represents the number of users whose preference to a AP falls in the corresponding ratio bins (the length of each bin is $0.1$) in the $x$ axis. We observe that there are about $2500$ users (the highest bar) with more than $90\%$ connections served by his most dominant AP. From the cumulative distribution, we observe that about $60\%$ users with $50\%$ connections served by his most dominant AP. Furthermore, we investigate how much each user's top-$K$ APs (ranked by the number of connections) contribute for his total requests in Fig.~\ref{fig:home_ap_topK}. We observe that there are more than $60\%$ (\emph{resp.} $90\%$) users with the majority (e.g., over $50\%$) of their connections served by their top $1$ (\emph{resp.} $5$) AP(s).

Given these observations, we claim that the AP set connected by each user shows an obvious stability, i.e., each user associates with familiar APs frequently and a few APs (e.g., the top $1$ AP) serve the majority, if not all, of the user's requests. In this paper, we focus on the case study for \emph{one user and one AP}, where each dominant AP only serves for one user and the prefetching strategies proposed later are only conduct on each user's dominant AP, for the following reasons: i) The contribution of the most dominant AP is significant. ii) In our dataset, less than $9\%$ users have the same dominant AP with others. Presumably, each user's most dominant AP is his home AP. We will investigate the collaborative content prefetching among multiple APs in our future works.

\subsection{User TV Series Watching Pattern} \label{section:tv_watching_pattern}

\begin{figure}[t]
 		\begin{minipage}[t]{0.3\linewidth}
	     \centering
	         \includegraphics[width=0.8\linewidth]{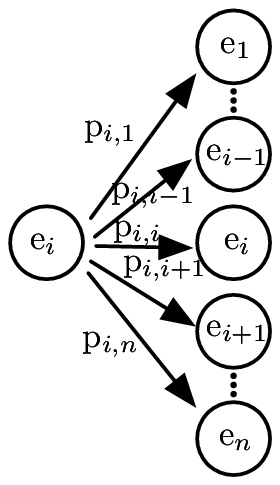}
			 \vspace{-0.2cm}
	     \caption{Possible transition among episodes in the same TV series.}
	      \label{fig:episode_transition}
 		\end{minipage}
 		 \hfill
 		\begin{minipage}[t]{0.63\linewidth}
	    \centering
	        \includegraphics[width=\linewidth]{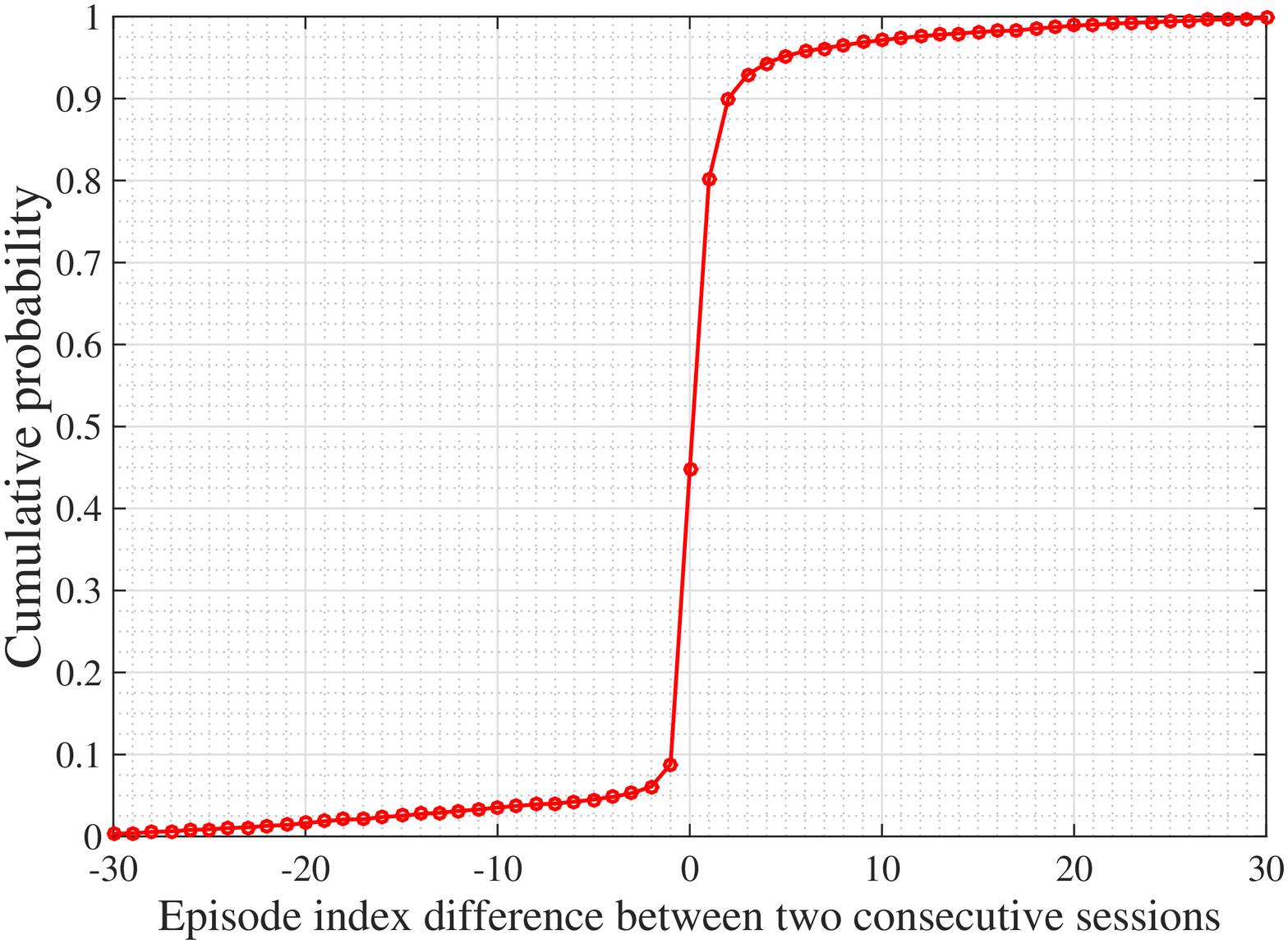}
			\vspace{-0.6cm}
	    \caption{Distribution of episode index difference between two consecutive sessions for a demonstrative TV series consisting of 33 episodes.}
     \label{fig:episode_transition_probability}
 		\end{minipage}
 \end{figure}

As illustrated in Fig.~\ref{fig:episode_transition}, after viewing the current episode, the user is likely to watch every episode of a TV series in consecutive time slots. By studying the video traces, we plot the transition probability distribution of the episodes index difference between two consecutive sessions in a demonstrative TV series in Fig.~\ref{fig:episode_transition_probability}. We observe that users are prone to either keep watching the current episode or the next three episodes, with a probability of $35\%$, $47\%$ respectively. The rationale is that there is strong relationship among the plot development of adjacent TV series. Besides, we observe that there is less possibility (about $18\%$) for large forward/backward episode transitions (large episode transitions are defined as the index differences range from $-30$ to $-1$ or from $4$ to $30$) and the cumulative possibility increases linearly with the large index differences. This is reasonable since the large episode transition indicates that the user does not care the plot development, as such, each episode is considered equally and the possibility for jumping to each episode is uniform.

These results motivate us to design the AP-assisted strategy to improve users' quality of experience (QoE) by prefetching more content while accounting for the fact that more prefetched content incur more additional monetary cost and resource competition costs. Details on the various cost models will be elaborated in the ensuing section (Sec.~\ref{sec:cost_model}).

\section{Model \& Formulation} \label{section:formulation}
In this section, we present some models based on the MDP framework~\cite{puterman2014markov} and formulate the AP-assisted content prefetching problem as an unconstrained optimization problem.

\subsection{System Architecture}
\begin{figure}[t]
    \centering
        \includegraphics[width=.8\linewidth]{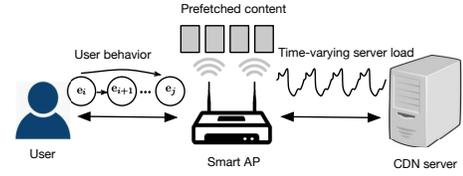}
    \caption{The architecture of AP-assisted content prefetching.}
     \label{fig:architecture}
\end{figure}

We illustrate the proposed system architecture in Fig.~\ref{fig:architecture}. The CDN server is the source of content and the server load is dynamic over time. Since the smart APs (e.g., HiWiFi\footnote{http://www.hiwifi.com.}, MiWiFi\footnote{http://www.miwifi.com.}, Newifi\footnote{http://www.newifi.com.}) are equipped with huge storage space, they are potential to help the content delivery. In this system, the distributed APs and the CDN servers are rent by the content service provider. Given the observation of time-varying server load, the content service provider should adopt the prefetching technique to shift traffic at peak periods, i.e., content should be prefetched to APs at server idle time slots. As such, when a request arrives, the AP first checks whether this request can be satisfied locally, i.e., whether the requested content has been stored in the AP. Otherwise, the request will be redirected to the remote CDN server, thereby incurring a higher delay. Meanwhile, the decision on which videos should be prefetched is driven by the users' watching behavior, i.e., the transition between episodes, to improve the possibility that the prefetched videos will be actually watched in the future.

\subsection{System Assumptions}

\subsubsection{Content Model} \label{section:content_model}
In this paper, we assume that the content catalog does not change in some periods (e.g., several hours or days). Especially, we focus on the on-demand TV series and assume the size of each TV episode is the same. Since the Time-to-Live (TTL) based caching policy: i) decouples the eviction mechanism among content~\cite{Berger:2014:EAT:2591971.2592038}; ii) captures the properties of existing popular eviction policies~\cite{fofack2012analysis} (e.g., LRU, FIFO and RND), we assume that the content stored in AP should be evicted after a time threshold $T_{th}$ is due, i.e., each content has a lifetime. Given the observation that the mean watching finish ratio of each episode is $72\%$~\cite{chen2013measurement}, we treat an episode as an unit and do not consider the partial prefetching. Furthermore, due to the facts: i) The end user's downlink bandwidth is limited; ii) Prefetching too many cotent items will lead to content eviction before being watching, we assume that at most $K_{th}$ videos can be prefetched during one time slot. 

\subsubsection{User Behavior Model}
We model users' transition behavior among series in the same TV series as a Markov model based on the insights learned from the measurement study in Sec.~\ref{section:tv_watching_pattern}. As opposed to the short videos published on social video sharing sites where users browser videos quickly~\cite{carlier2015video}, users tend to complete watching episodes of TV series~\cite{chen2013measurement}. We set the duration of each time slot the same value for each episode, thus the user only watches one video in one time slot. Since the prediction of periods when users will watch videos is outside the scope of the paper, our design works in a conservative way, i.e., our design only works in \emph{effective time slots}, which are defined as the time slots when a user is actually watching videos and we do not consider content prefetching during time slots when he does not consume videos.

\begin{figure}[t]
    \centering
        \includegraphics[width=\linewidth]{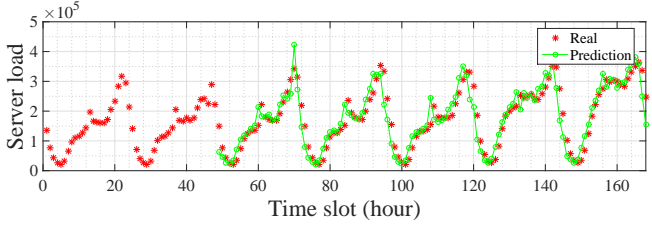}
    \caption{CDN server load over time measured using traces from The Video Service Provider.}
     \label{fig:serverload}
\end{figure}

\subsubsection{Network Traffic Model}\label{section:network_traffic_model}

We study the number of video requests hourly from the same dataset as we used in Sec. 2, to build the network traffic model. From Fig.~\ref{fig:serverload}, we observe the marked daily pattern of the video request traces, which motivates us to adopt the seasonal ARIMA model (Autoregressive Integrated Moving Average)~\cite{mills1990time} to predict the server load. Mathematically, 
\begin{equation}
	\Phi(L^{s}) \phi(L) \nabla_{s}^{D} \nabla^{d} X_{t} = \theta(L) \Theta(L^{s})\varepsilon_{t},
\end{equation}
where $\Phi$ (\emph{resp.} $\phi$) and $\Theta$ (\emph{resp.} $\theta$) are the seasonal (\emph{resp.} non-seasonal) autoregressive and moving average parts. $\varepsilon_{t}$ is the white noise for the stationary distribution. $L$ is the lag operation, i.e., $L^{d}X_{t} = X_{t-d}$. $\nabla$ is the difference operator, i.e., $\nabla X_{t}= X_{t} - X_{t-1}$ and thus, $\nabla = 1 - L$. $D$ (\emph{resp.} $d$) is the the order of seasonal (\emph{resp.} non-seasonal) integration part and $s$ denotes the number of periods in a season. Due to the limit of space, we omit the details of ARIMA model which can be found in~\cite{mills1990time}.

By training with the historical traces, the model is of the form of $(0,1,1) \times (0,1,1)_{24}$ and there are $24$ periods, one for every hour of the day, in a season. The prediction results are presented in Fig.~\ref{fig:serverload}, each green dot is predicted by learning the server load in previous $48$ hours. We observe that the prediction achieves a relatively accurate estimation (e.g., Mean Absolute Percent Error is $17.52\%$) of the server load with a small learning window.

\subsubsection{Cost Model}\label{sec:cost_model}
The costs for content prefetching consist of several parts, including: the transmission cost for prefetching the content from the server to the AP or directly downloading to the end user; the storage cost for holding the content in the AP; the QoE degradation cost for the increased delay by fetching the content from the server rather than the AP and the resource competition cost for the higher startup delay incurred by limited resource competition. In particular, the QoE degradation cost and resource competition cost are designed to reward the system for the improvement of user-perceived quality of experience.

\textbf{Transmission Cost:} By taking the server load, which can be predicted with the network traffic model in Sec.~\ref{section:network_traffic_model}, into consideration, we adopt a server load aware transmission cost function $\Psi(l)$. It is worth noting that there is no requirement for the exact definition of $\Psi(l)$ as long as it is a increasing convex function. Here, we follow the work in~\cite{zheng2012utilization} and adopt the logarithmic barrier function as follows.
\begin{equation}
  \Psi(l) = - log(1-\frac{l}{l_{th}}),
\end{equation}
where $l$ is the current server load and $l_{th}$ is constant variable which denotes the server threshold load. Therefore, $\Psi(l)$ is a strictly increasing convex function with respect to the server load. The rationale behind $\Psi(l)$ is that it is cheap to prefetch content when the server is under small utilization whereas when the load approaches the threshold load $l_{th}$, we get highly penalized to guarantee that content prefetching should never happen.

Thereby, the cost for content transmission is defined as follows.
\begin{equation}
	C^{tr}(x) = \beta \Psi(l) x,
\end{equation}
where $x$ is the number of content items to be downloaded/prefetched from the CDN server and $\beta$ is the tuning parameter to guarantee that the median of the transmission cost per content is consistent with the Amazon on-demand pricing model~\cite{amazon_price} and the details will be elaborated in Sec.~\ref{section:parameter}.

\textbf{Storage Cost:} The cost for storing content in AP for one time slot is defined as follows.
\begin{equation}
	C^{st}(x) = \kappa x,
\end{equation}
where $\kappa$ is the fixed cost per content per time slot in the AP, $x$ is the number of stored content items.

\textbf{Latency Increase Cost:} We also consider the cost introduced by the QoE degradation due to cache miss on AP, as follows.
\begin{equation}
	C^{la}(x) = (d^1 - d^0) x,
\end{equation}
where $x$ is the number of content items to be downloaded from the CDN server, $d^0$ is a fixed variable denoting the startup delay for downloading the content from the AP, while $d^1$ is a volatile variable denoting the startup delay for downloading the content from the server and its value depends on the current server load.

\textbf{Resource Competition Cost:} Since we schedule content prefetching while the user is watching a video, the prefetching task competes for the limited resources, e.g., the end users' downlink bandwidth, with the downloading task for the current playback. The competition incurs that the assigned bandwidth for the playback task decreases. Based on the relationship between QoE and the bandwidth in~\cite{chen2006quantifying}, we define the competition cost for content prefetching as follows.
\begin{equation}
		C^{p}(x, y) = \begin{cases}
		0 & x = 0;\\
		0 & y = 0; \\
		log(bw) - log(\frac{bw}{x+y}) = log(x+y) & otherwise,
		\end{cases}
\end{equation}
where $bw$ is the residential downlink bandwidth, $x$ is the number of content items to be prefetched in parallel with the video playback and $y$ is a binary variable, i.e., $y=0$ if the being watched video has been previously prefetched; otherwise, $y=1$. It is reasonable to expect that the resource competition cost is zero either when there are no prefetching tasks or the video being watched has been previously prefetched. Note that, for simplicity, we consider that the bandwidth is allocated among the tasks equally.

\subsection{Problem Formulation} \label{sec:formulation}

\subsubsection{System States}
Let $\mathcal{S}=\{s_{1}, s_{2}, ..., s_{t}, ..., s_{T}\}$ denote the state space. $s_{t} = (e_{t}^{w}, \textbf{E}_{t}^{st})$, where $e_{t}^{w}$ is the content actually being watched by the user at time slot $t$ and $\textbf{E}_{t}^{st}$ is the content set whose elements have been stored in the AP. Therefore, each state can represent both the episode user is watching and the content set stored in the AP. 

\subsubsection{Actions}
Let $\mathcal{A} = \{a_{1}, a_{2}, ..., a_{t}, ..., a_{T}\}$ denote the action set adopted by the system over time. $a_{t} = (\textbf{E}_t^{tr}, \textbf{E}_t^{d})$, where $\textbf{E}_t^{tr}$ and $\textbf{E}_t^{d}$ are the content set scheduled to be prefetched and deleted at time slot $t$, respectively. Note that no action is taken for prefetching if the size of $\textbf{E}_t^{tr}$ is $0$ and the elements of $\textbf{E}_t^{d}$ are determined by the content lifetime based on the eviction policy presented in Sec.~\ref{section:content_model}.

\subsubsection{System State Transition}
Intuitively, the system state in time slot $t+1$ is $s_{t+1} = (e_{t+1}^{w}, \textbf{E}_{t+1}^{st})$, where $\textbf{E}_{t+1}^{st}$ can be calculated based on its previous state at time slot $t$, as follows.
\begin{equation}
 \textbf{E}_{t+1}^{st}=\textbf{E}_t^{st} \cup \textbf{E}_t^{tr} \setminus \textbf{E}_t^{d}.
\end{equation} 

The transition possibility from $s_t$ to $s_{t+1}$ can be calculated as follows.
\begin{equation} \label{eq:possibility}
	\begin{split}
	P_{a_{t}}(s_{t}, s_{t+1}) &= P(s_{t+1} \mid s_{t}, a_{t}) \\
		 					&= P((e_{t+1}^{w}, \textbf{E}_{t+1}^{st}) \mid (e_{t}^{w}, \textbf{E}_{t}^{st}), a_{t} )\\
							&=P(e_{t+1}^{w} \mid e_{t}^{w}) P(\textbf{E}_{t+1}^{st} \mid \textbf{E}_{t}^{st}, a_{t}).
	\end{split}
\end{equation}
The last equality in Eq.~(\ref{eq:possibility}) follows because the possibility of episodes' transition is independent of the action.

\subsubsection{Cost Function} \label{section:cost}

The cost function at time slot $t$ is defined as the weighted sum of the aforementioned parts of costs (Eq.~\ref{eq:cost_function}), including the monetary-related cost part (Eq.~\ref{eq:monetary_function}) and the QoE-related cost part (Eq.~\ref{eq:qoe_function}). 
\begin{equation}\label{eq:cost_function}
  g_{t}(s_t, a_t) = C^{m} + \lambda_{1} C^{q},
\end{equation}
\begin{equation}\label{eq:monetary_function}
  C^{m} = C^{tr}(\card{\textbf{E}_t^{tr}}) + C^{tr}(\delta(e_{t}^{w}, \textbf{E}_t^{st})) + C^{st}(\card{\textbf{E}_t^{st} \cup \textbf{E}_t^{tr}}),
\end{equation}
\begin{equation}\label{eq:qoe_function}
  C^{q} = C^{la}(\delta(e_{t}^{w}, \textbf{E}_t^{st})) + \lambda_{2} C^{p}(\card{\textbf{E}_t^{tr}}, \delta(e_{t}^{w}, \textbf{E}_t^{st})),
\end{equation}
where $\lambda_{1}$ is a positive trade-off parameter to balance the trade-off between the $C^{m}$ and $C^{q}$, $\lambda_{2}$ is a positive tuning parameter for balancing the relationship between latency cost $C^{la}$ and resource competition cost $C^{p}$, $\card{\cdot}$ is the operator of calculating the number of elements in the argument set \text{``$\cdot$''} and $\delta(e_{t}^{w}, \textbf{E}_t^{st})$ is an indicator function defined as follows.
\begin{equation}
   \delta(e_{t}^{w}, \textbf{E}_t^{st}) = \begin{cases}
     0 & e_{t}^{w} \in  \textbf{E}_t^{st};\\
     1 & otherwise.
   \end{cases} 
 \end{equation}

Our objective is to minimize the total cost over a finite time horizon, i.e.,
\begin{equation}
	\min\limits_{\{a_{t}\}} \sum_{t=0}^{T-1} g_{t}(s_t, a_t).
\end{equation}

\section{Strategies for AP-assisted Content Prefetch} \label{section:algorithm}
In this section, we first present an illustrative example to elaborate details of the problem. Then we start with some naive algorithms, including a random fixed algorithm for the lower performance bound, and a backward induction algorithm for the upper performance bound. Besides, we also propose a heuristic algorithm as another baseline. Finally, we adopt the reinforcement learning algorithm to design an online strategy, which is more practical.

\begin{figure}[t]
    \centering
        \includegraphics[width=\linewidth]{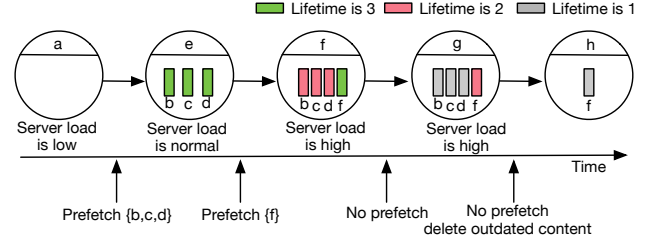}
    \caption{Illustration of the content prefetching according to the user behavior and the time-varying server load.}
     \label{fig:state_transition_case}
\end{figure}

We illustrate the main idea behind our design by a simple example in Fig.~\ref{fig:state_transition_case}. In this case, the users' video watching sequence is $a \rightarrow e \rightarrow f \rightarrow g \rightarrow h$ and the $T_{th}$ is $3$, i.e., a content should be evicted after $3$ time slots. Since the server load is low when user is watching video $a$, it is optimal to prefetch many content based on the transition probability. Less or no content should be prefetched when the server is overloaded. Note that in the last state transition, content $b, c, d$ are evicted because of time out. Intuitively, the optimization objective is to make sequential optimal decisions on the content prefetching over a finite time horizon.

\subsection{Random Algorithm \& Performance Lower Bound}

In order to obtain the lower performance bound, we consider the naive strategy which prefetches a fixed number of content and the content are randomly selected. Since the system has no idea of the user behavior and the server load information in the future, how many and what content should be prefetched are hard to decide. As such, all the choices within the prefetch threshold $K_{th}$, i.e., no prefetching, one episode prefetching, and so forth, are enumerated and a fixed number of content are randomly selected from the episodes set of a TV series. As for each choice, the total cost is calculated as the sum of cost at each timeslot (Eq.~(\ref{eq:cost_function})). Recall that the cost at each timeslot is calculated as the weighted sum of the monetary-related cost and the QoE-related cost. After the traversal, the average cost of all choices is calculated as the cost expectation.

\subsection{Heuristic Algorithm}
We also propose a heuristic algorithm for performance comparison. As opposed to the random algorithm, the heuristic algorithm incorporates the users' TV watching pattern studied in Sec.~\ref{section:tv_watching_pattern} into the strategy design. Since users are prone to either keep watching the current episode or the next three episodes, we choose to prefetch content from the next three episodes. However, the next period when the user will watch videos is unknown beforehand, thus we can not decide the optimal number of content to be prefetched based on the varying server loads at the current period and next period. In this algorithm, we enumerate the possible choices on the number of prefetching and calculate the average cost of all choices as the cost expectation.

\subsection{Offline Algorithm \& Performance Upper Bound}
In order to obtain the upper performance bound, we assume the complete user behavior trace $\mathcal{H}=\{e_{1}^{w}, \ldots, e_{T}^{w}\}$ is known in prior, which means that not only the information before current time but also the future information, e.g., what and when the user will watch, are available. Thus, we can design the offline algorithm which achieves the optimal performance. Given this assumption, the original problem reduces to a deterministic MDP and the transition possibility can be calculated as follows.
\begin{equation} \label{eq:determine_transfer}
    \widetilde{P}_{a_{t}}(s,s') = \begin{cases}
       1 & e^{w}=e_{t}^{w}, e^{w'}=e_{t+1}^{w}, \textbf{E}^{st'}=\textbf{E}_t^{st} \cup \textbf{E}_t^{tr} \setminus \textbf{E}_t^{d};\\
       0 & otherwise.
    \end{cases}
\end{equation}

Since the Bellman optimality equation~\cite{bellman1952theory} characterizes the value function and calculates the optimal value in state $s$ as the sum of the immediate cost and the discounted optimal value in the next state, we define the value function of taking action $a_t$ in state $s_t$ as $\textbf{V}_{t}(s_t, a_t)$, which represents the expected accumulative cost from the current time slot $t$ to the end $T$. 
\begin{equation} \label{eq:value_function}
	\textbf{V}_{t}(s_t, a_t) = g_{t}(s_t, a_t) + \sum\limits_{s_{t+1} \in \mathcal{S}} \widetilde{P}_{a_{t}}(s_{t}, s_{t+1}) \textbf{V}_{t+1}^{\ast}(s_{t+1}).
\end{equation}

The minimal value function $\textbf{V}_{t}^{\ast}(s_t)$ can be derived as follows. Note that we initially set $\textbf{V}_{T}^{\ast}(s_{T}) = 0$ for all $s_{T} \in \mathcal{S}$. 
\begin{equation} \label{eq:optimal_value_function}
	\textbf{V}_{t}^{\ast}(s_t) = \min\limits_{a_t \in \mathcal{A}} \textbf{V}_{t}(s_t, a_t).
\end{equation}

After finding the optimal value of $\textbf{V}_{t}(s_t, a_t)$, we can derive the optimal policy using the following equation.
\begin{equation}\label{eq:optimal_policy}
	\pi(s_t, t) = \argmin_{a_t \in \mathcal{A}} \textbf{V}_{t}(s_t, a_t),
\end{equation}
where $\pi$ is the policy which maps the state and stage to actions, i.e., $\pi: s \times t \to a$.

The details of this method are presented in Algorithm~\ref{algorithm:offline_algorithm}. It adopts the value iteration technique, which utilizes the Bellman optimality equation, to iteratively compute the expected minimal accumulated cost at each time slot. As such, the optimal policy at each time slot can be derived. In particular, the value iteration begins at the end time slot $T$ with an initial value, e.g., $0$, for the value function $\textbf{V}_{T}^{\ast}(s_T)$ (line~\ref{line:initialization}) and works backward, calculating the value at each time slot (line~\ref{line:backward_begin}$-$line~\ref{line:backward_end}). After that, the optimal policy is derived forward based on the stored values (line~\ref{line:forward_begin}$-$line~\ref{line:forward_end}). 

\IncMargin{1em}
\begin{algorithm}[t]
  
    \DontPrintSemicolon
	\SetKwInOut{Input}{Input}
	\SetKwInOut{Output}{Output}
	
	\Indm
    \Input{a complete user behavior traces $\mathcal{H}$ for a TV series} 
    \Output{optimal policy $\pi(s_t, t)$ at each time slot}
	\BlankLine
	\Indp
	 initialize the value function $\textbf{V}_{T}^{\ast}(s_T)=0$ for all $s_{T} \in \mathcal{S}$\;\label{line:initialization}
	\For {$t = T - 1, \ldots, 1$} {\label{line:backward_begin}
			\For {each $a_t \in \mathcal{A}$} {
		    calculate the transition possibility $\widetilde{P}_{a_{t}}(s_{t}, s')$ with Eq.~(\ref{eq:determine_transfer})\;
			update the value function $\textbf{V}_{t}(s_t, a_t)$ with Eq.~(\ref{eq:value_function})\;
			}
			derive the minimal value function $\textbf{V}_{t}^{\ast}(s_t, a_t)$ with Eq.~(\ref{eq:optimal_value_function})\;
		}\label{line:backward_end}
	\For {$t = 1, \ldots, T - 1$} {\label{line:forward_begin}
		derive the optimal policy $\pi(s_t, t)$ with Eq.~(\ref{eq:optimal_policy})\;
		}\label{line:forward_end}
	\caption{Offline Algorithm for MDP.}\label{algorithm:offline_algorithm}	
\end{algorithm}
\DecMargin{1em}

\subsection{Online Algorithm with Approximate Reinforcement Learning}
In this part, we propose an online algorithm using reinforcement learning. Firstly, the problem of the large state and action spaces for our content prefetching paradigm is analyzed. Then, we propose to use the function approximation technique, i.e., the use of a parameterized functional form to represent the value function.

\subsubsection{Curse of Dimensionality}
Based on the definition of state and action variable in Sec.~\ref{sec:formulation}, we analyze the dimensionality of state space and the action space. Recall that $s_{t} = (e_{t}^{w}, \textbf{E}_{t}^{st})$ consists of two parts: the current watching episode and the buffered content set. As for a TV-series with $m$ episodes and each prefetched content with $T_{th}$ lifetime, there are $\card{\mathcal{S}} = m (T_{th} + 1)^m$ states. Similarly, the action variable $a_{t} = (\textbf{E}_t^{tr}, \textbf{E}_t^{d})$ consists two parts: the content to be prefetched and the content to be deleted. The number of content to be prefetched is smaller than the threshold $K_{th}$ and the deletion action is enforced by the lifetime threshold ($T_{th}$ time slots). Hence, there are $\card{\mathcal{A}} = \binom m1 + \binom m2 + \ldots + \binom m{K_{th}}$ actions for each state, where the first term indicates a random prefetching on just one episode, the second term indicates a random prefetching on two episodes, and so forth. 
% Unsurprisingly, there are far too many states and actions.
As for our problem, the classic \emph{tabular} method, which represents the value function as a table with an
entry for each state or state-action pair, could not be adopted. Because the large size of the table ($\card{\mathcal{S}}\times\card{\mathcal{A}}$) requires many memory and much time to accurately calculate them.

\subsubsection{Gradient-based Q-learning}
Approximation-based value function for problems with large space is an active research topic on reinforcement learning~\cite{sutton1998reinforcement}, which adopts approximate functions to generalize the value of states. And we will reduce the state and action dimensionality by feature extraction and a heuristic policy, respectively.

\textbf{Compact Parameterized Function Representation:} As for the state space, we adopt a feature-extraction function $\phi: \mathcal{S} \to \Phi$ to map states into features in the feature space $\Phi$. Corresponding to each state $s$, there is a feature vector $\textbf{x}(s)$, $\phi(s) = \textbf{x}(s) = (x_{1}(s), x_{2}(s), \ldots, x_{n}(s))^{\top}$. In our problem, we represent an episode with a $(T_{th} + 2)$ binary vector: i) The first component is 1 if the episode is being watched; otherwise, it is 0. ii) The $(i+1)$-th component of the binary vector is 1 if the remaining lifetime of the episode is $i$; otherwise, it is 0. Hence, each state can be represented with a binary vector of size $m(T_{th} + 2)$. After feature extraction, the approximated $Q$ function is defined as follows.
\begin{equation}\label{eq:q_function}
	Q_{t}(s_t, a_t) = \boldsymbol\theta_{t}(a_t)\phi(s_t)=\boldsymbol\theta_{t}(a_t)\textbf{x}(s_t),
\end{equation}
where $\boldsymbol\theta_{t}$ denotes the adaptable parameter matrix of size $\card{\mathcal{A'}} \times \card{\mathcal{S'}}$ at time $t$ and $\boldsymbol\theta_{t}(a_t)$ denotes $a_{t}$-th row of $\boldsymbol\theta_{t}$. $\mathcal{A'}$ and $\mathcal{S'}$ denotes the reduced action space and state space, respectively.
 
The dimensionality of the action space is reduced by the insights learned in Sec.~\ref{section:tv_watching_pattern}. Since the probability that user jumps to the $(i+4)$-th episodes after watching $i$-th is too small, we only consider three classes of actions: prefetch one, two or three episodes from next consecutive three episodes. And these classes have $3$, $3$, $1$ choices, respectively.

By representing in the compact way, at each time slot, we only need to learn the parameter matrix $\boldsymbol\theta_{t}$. In our problem, $\card{\mathcal{A'}} = 7$, $\card{\mathcal{S'}}=m(T_{th} + 2)$, and the size of $\boldsymbol\theta_{t}$ is $\card{\mathcal{A'}} \times \card{\mathcal{S'}}=7 \times m(T_{th} + 2)$, which is very manageable.

To summarize, we dramatically reduce the dimensionality from $\card{\mathcal{A}} \times \card{\mathcal{S}}$ to $\card{\mathcal{A'}} \times \card{\mathcal{S'}}$.

\textbf{Gradient-descent Update:} We quantify the one step temporal-difference error as $f(s_t, a_t) = \frac{1}{2} (\delta_{t})^2$, where $\delta_{t}=\gamma \min\limits_{a} Q_t(s_{t+1}, a) + g_{t}(s_t, a_t) - Q_t(s_t, a_t)$, and $\gamma$ $(0 \leq \gamma \leq 1)$ is the discount factor, which can be interpreted intuitively as a way of trading off the importance of sooner and later cost.

In order to minimize the temporal-difference error, we first calculate the negative gradient of $f(s_t, a_t)$ as follows. 
\begin{equation}\label{eq:gradient}
	\begin{split}
	-\nabla_{\boldsymbol\theta_{t}} f(s_t, a_t) &= -\delta_{t}\nabla_{\boldsymbol\theta_{t}} (\delta_{t})\\
												&=\delta_{t}\nabla_{\boldsymbol\theta_{t}} Q_{t}(s_t, a_t).
	\end{split}
\end{equation}

Then, we update the parameters with Q-learning along the negative gradient descent direction, in which the error falls most rapidly, as follows.
\begin{equation}\label{eq:parameter_update}
	\begin{split}
\boldsymbol\theta_{t+1}(a_t) &= \boldsymbol\theta_t(a_t) - \alpha \nabla_{\boldsymbol\theta_{t}} f(s_t, a_t) \\
							 &= \boldsymbol\theta_t(a_t) + \alpha\delta_{t} \nabla_{\boldsymbol\theta_{t}} Q_{t}(s_t, a_t)\\
							 &= \boldsymbol\theta_t(a_t) + \alpha\delta_{t}\phi(s_t),
	\end{split}
\end{equation}
where $\alpha_{t}$ $(0 < \alpha \leq 1)$ is a step-size parameter, which influences the step learning rate, i.e., how much the newly acquired information will override the old information.

\textbf{Exploration-exploitation Balance:} In order to guarantee that the Q-learning converges to the optimal Q-function, we adopt the $\epsilon$-greedy policy to balance the exploration of selecting any action with a non-zero probability with exploitation of selecting the greedy actions in the current Q-function. In particular, the action is selected according to
\begin{equation}\label{eq:greedy_policy}
    \pi(s_t,t) = \begin{cases}
      \argmin\limits_{a \in \mathcal{A'}} (\boldsymbol\theta_{t}(a)\textbf{x}(s_t)) & \text{with probability} (1-\epsilon);\\
       \text{random action from } \mathcal{A'} & \text{with probability } \epsilon,
    \end{cases}
\end{equation}  
where $\epsilon$ ($0 \leq \epsilon < 1$) is the exploration probability at state $s_t$ and $\epsilon$ diminishes over time with a rate $\epsilon'$.

\textbf{Algorithm Design:} In order to speed up the learning process, we extend the Q-learning with the experience replay~\cite{lin1992self}. The details of the online approach are presented in Algorithm~\ref{algorithm:experience_replay} and Algorithm~\ref{algorithm:online_algorithm}. 

During the experience replay process (Algorithm~\ref{algorithm:experience_replay}), we first extract the feature vectors $\textbf{x}(s_t)$ for each state (line~\ref{line:feature_extraction}) and select the action based on the $\epsilon$-greedy policy. After performing the action, the parameter matrix $\boldsymbol\theta_{t}$ is updated based on the one step temporal-difference error (line~\ref{line:parameter_update}). This process continues until the parameter matrix $\boldsymbol\theta_{t}$ begins to converge. Otherwise, more traces will be collected for the experience replay. The replay process is guaranteed to converge when the MDP problem is finite~\cite{watkins1992q}.

After obtaining the converged parameter matrix $\boldsymbol\theta_{t}$ at each time slot, we can derive the Q-value for any state at the current time slot. As illustrated in Algorithm~\ref{algorithm:online_algorithm}, the policy is derived by choosing the action with minimum Q-value $Q_{t}(s_t, a_t)$ (line~\ref{line:policy_derivation}). Note that the learning process continues for a better policy derivation in the future (line~\ref{line:online_update}).

\IncMargin{1em}
\begin{algorithm}[t]
  
    \DontPrintSemicolon
	\SetKwInOut{Input}{Input}
	\SetKwInOut{Output}{Output}
	
	\Indm
    \Input{discount factor $\gamma$, learning rate $\alpha$, feature-extraction function $\phi$, greedy-$\epsilon$ policy, user behavior history $\mathcal{H} = \{e_{1}^{w}, \ldots, e_{T'}^{w}\}$ and the convergence threshold $\boldsymbol\chi$} 
    \Output{the parameter matrix $\boldsymbol\theta$}
	\BlankLine
	\Indp
	 initialize the parameter matrix $\boldsymbol\theta_{1}$ (e.g., 0)\;
	\For {$t = 1, \ldots, T' - 1$} {
			map $s_t$ to a feature vector $\textbf{x}(s)=\phi(s)$\;\label{line:feature_extraction}
			apply $a_t=\pi(s_t,t)$ with Eq.~(\ref{eq:greedy_policy})\;
			measure the next state $s_{t+1}$ and the cost $g_{t}(s_t, a_{t})$\;
			update the parameter $\boldsymbol\theta_{t}$ with Eq.~(\ref{eq:parameter_update})\;\label{line:parameter_update}
			$\epsilon = \epsilon - \epsilon'$
		}
	\eIf {any $|\boldsymbol\theta_{t} - \boldsymbol\theta'_{t}| \geq \boldsymbol\chi$} {
		collect more user behavior history
	}{
	\KwRet parameter matrix $\boldsymbol\theta$
	}
	\caption{Q-learning with Experience Replay.}\label{algorithm:experience_replay}	
\end{algorithm}
\DecMargin{1em}

\IncMargin{1em}
\begin{algorithm}[t]
  
    \DontPrintSemicolon
	\SetKwInOut{Input}{Input}
	\SetKwInOut{Output}{Output}
	
	\Indm
    \Input{discount factor $\gamma$, learning rate $\alpha$, feature-extraction function $\phi$, state $s_t$ at time slot $t$, and the learned parameter matrix $\boldsymbol\theta_{t}$ at each time slot from experience replay process}
    \Output{optimal policy $\pi(s_t, t)$ at each time slot}
	\BlankLine
	\Indp
	 
	\For {$t = 1, \ldots, T-1$} {
			map $s_t$ to a feature vector $\textbf{x}(s)=\phi(s)$\;
			\For {each $a_t \in \mathcal{A'}$} {
			update the approximated $Q$ function $Q_{t}(s_t, a_t)$ with Eq.~(\ref{eq:q_function})\;
			}
			apply $\pi(s_t,t) = \argmin\limits_{a \in \mathcal{A'}} Q_{t}(s_t, a_t)$\;\label{line:policy_derivation}
			measure the next state $s_{t+1}$ and the cost $g_{t}(s_t, a)$\;
			update the parameter $\boldsymbol\theta_{t}$ with Eq.~(\ref{eq:parameter_update})\;\label{line:online_update}
		}
	\caption{Online Learning with Function Approximation.}\label{algorithm:online_algorithm}	
\end{algorithm}
\DecMargin{1em}
\section{Performance Evaluation}\label{section:evaluation}
In this section, we numerically evaluate the effectiveness and performance of our AP-assisted content prefetching framework based on trace-driven simulations. Specifically, we use the real traces from The Video Service Provider to characterize users' watching patterns.
\subsection{Experiment Setup}
\subsubsection{Parameters Setting}\label{section:parameter}
According to the Amazon's on-demand model~\cite{amazon_price}, i.e., the storage cost is $2 \times 10^{-4}$ USD/GB per hour and the transmission cost is $0.12$ USD/GB, we calculate the corresponding cost parameters in our cost model as follows: i) The storage cost per content item per time slot $\kappa$ equals $6 \times 10^{-5}$. Note that we consider that all the episodes are high definition (HD videos), i.e., 720p video settings, and the duration of each episode is $45$ minutes, thus the size of each episode is $400$ MB. ii) We adjust the tuning parameter $\beta$ as $0.16$ in the cost model to guarantee the median of transmission cost for a content is $0.048$ USD, which is consistent with the pricing model~\cite{amazon_price} widely adopted by the literature~\cite{he2014cost}. iii) Since users start to abandon the video if the startup delay exceeds about $2$ s~\cite{krishnan2013video}, the video startup delay from the CDN server $d^{1}$ is less than $2$ s and is proportional to the degree of the server load. The video startup delay from AP $d^{0}$ equals $0.05$ s. iv) For Q-learning settings, we set the learning rate $\alpha = 0.5$ to give equal weight to new and old knowledge, the discount factor $\gamma=0.99$ to take more future costs into account, and the Q-value converage threshold $\boldsymbol\chi = 0.0001$. We initially set $\epsilon=0.5$ and the reduction rate $\epsilon'=0.0005$ per time slot.

Since content prefetching will incur competition cost at the current time slot and may reduce the startup delay in the next time slot, the tuning parameter $\lambda_{2}$ for balancing the relationship between latency cost $C^t$ and resource competition cost $C^p$ must be less than $1$. Otherwise, it will never be optimal to prefetch any content, because the benefit of content prefetching, i.e., the reduced startup delay, will be offset by the incurred competition cost. In the following experiments, if not otherwise specified, $\lambda_{2}$ is $0.02$ and the trade-off parameter between the monetary-related cost and the QoE-related cost ($\lambda_{1}$) is $0.9$. The details of experiment setting are summarized in Table~\ref{tab:parameter}.

\begin{table}[t]
    \centering
	\caption{Simulation Parameters Setting.} \label{tab:parameter}
	\ra{0.9}
    \begin{tabular}{p{0.6\linewidth} p{0.25\linewidth}<{\centering}}
        \toprule
        Parameter  &  Value \\
		\midrule
        Lifetime threshold $T_{th}$          &   $3$ time slots\\
        Prefetch content threshold at each time slot $K_{th}$     &   $3$ \\
	    Duration of one time slot              & $45$ min\\
	    Storage cost per content item per time slot $\kappa$              & $6 \times 10^{-5}$ \$\\
		Tuning parameter for transmission cost per content item $\beta$              & $0.16$\\
		Startup delay from CDN server $d^{1}$              & $2 \times \frac{l}{l_{th}}$ s\\
		Startup delay from AP $d^{0}$              & $0.05$ s\\
		% Resource competition cost per content $\zeta$              & $2 \times \frac{l}{l_{th}}$\\
        Discount factor of online learning $\gamma$              &   $0.99$\\
        Learning rate of online learning $\alpha$              & $0.5$ \\
		Convergence threshold $\boldsymbol\chi$				  &  $0.0001$ \\
        \bottomrule
    \end{tabular}

\end{table}

\subsubsection{Trace-driven Simulation}
According to the video session traces from The Video Service Provider, we simulate each user's behavior based on the request patterns and simulate the server load based on the overall requests from users at each time slot. Since the collected traces spanning limited periods, i.e., $2$ weeks, we focus on the TV series with $30$ episodes and remove users who do not issue more than $30$ requests for any TV series. It is worth noting that because the user may keep watching one episode in several consecutive time slots, $30$ requests for a TV series do not indicate the user watches $30$ distinct episodes. The performance of each algorithm presented in the following section is the average results of $1000$ rounds.

\subsubsection{Metrics}
In order to quantify the performance of different prefetching strategies, we adopt the following metrics: i) Precision ratio ($PR$)~\cite{domenech2006web}: the ratio between the number of the useful prefetches $p_{u}$ and the total number of prefetched videos $p_{t}$. Mathematically, $PR = \frac{p_{u}}{p_{t}}$. Here, $p_{u}$ is defined as the number of prefetched videos which are actually consumed before eviction. ii) Hit ratio ($HR$)~\cite{domenech2006web}: the ratio between the useful prefetches $p_{u}$ and the total number of requested videos $r_{t}$. Mathematically, $HR = \frac{p_{u}}{r_{t}}$. iii) Various costs formulated in Sec.~\ref{section:cost}, including the overall cost $g_{t}(s_t, a_t)$, the monetary-related cost $C^{m}$ and the QoE-related cost $C^{q}$.

\subsection{Experiment Results}
\subsubsection{Effectiveness of Our Proposed Strategies}
\begin{figure}[t]
     \begin{minipage}[t]{.46\linewidth}
          \centering
               \includegraphics[width=\linewidth]{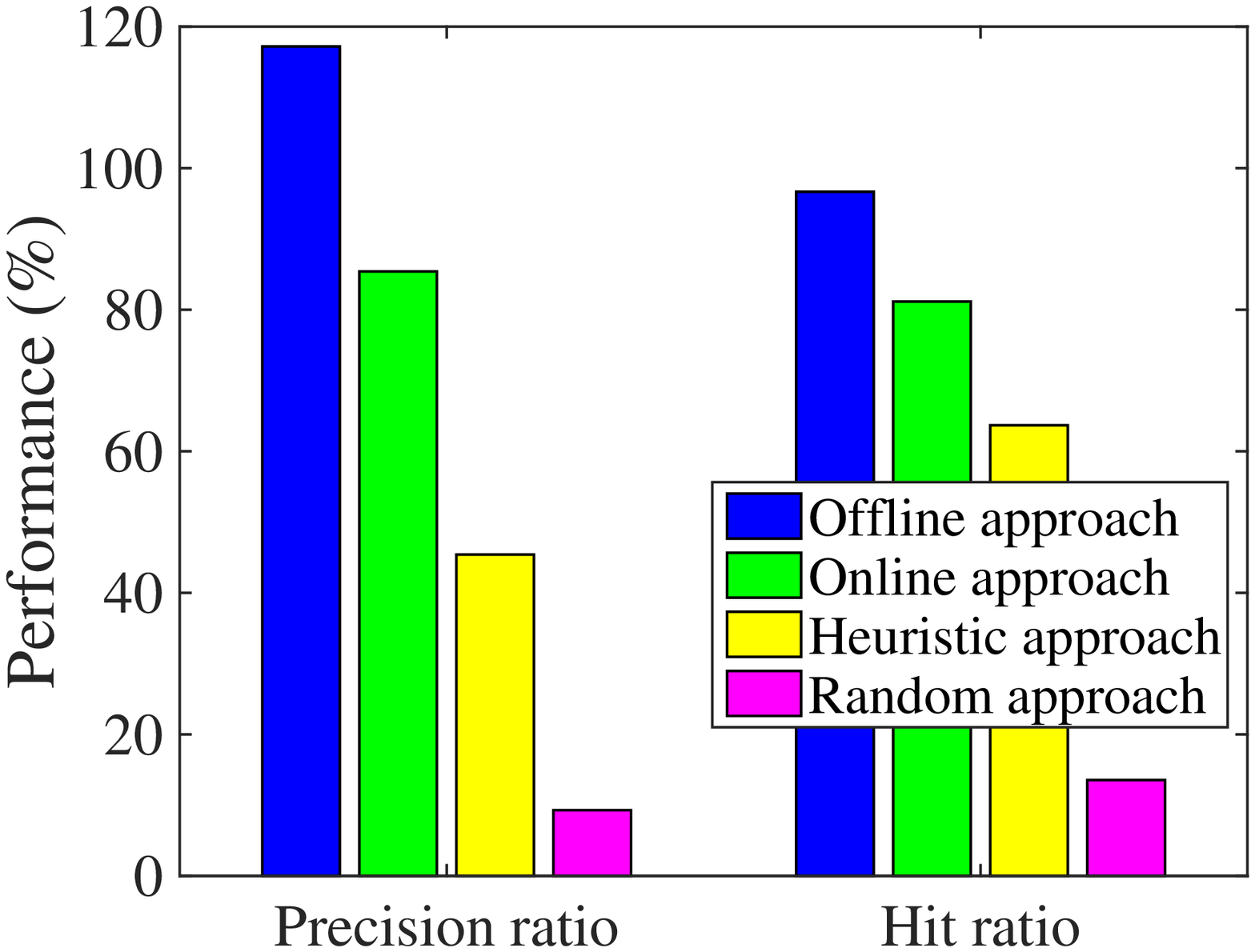}
		  \raggedright
			   \scriptsize (a) Precision and hit ratio with different approaches.
     \end{minipage}
     \hfill
     \begin{minipage}[t]{.46\linewidth}
          \centering
			   \includegraphics[width=\linewidth]{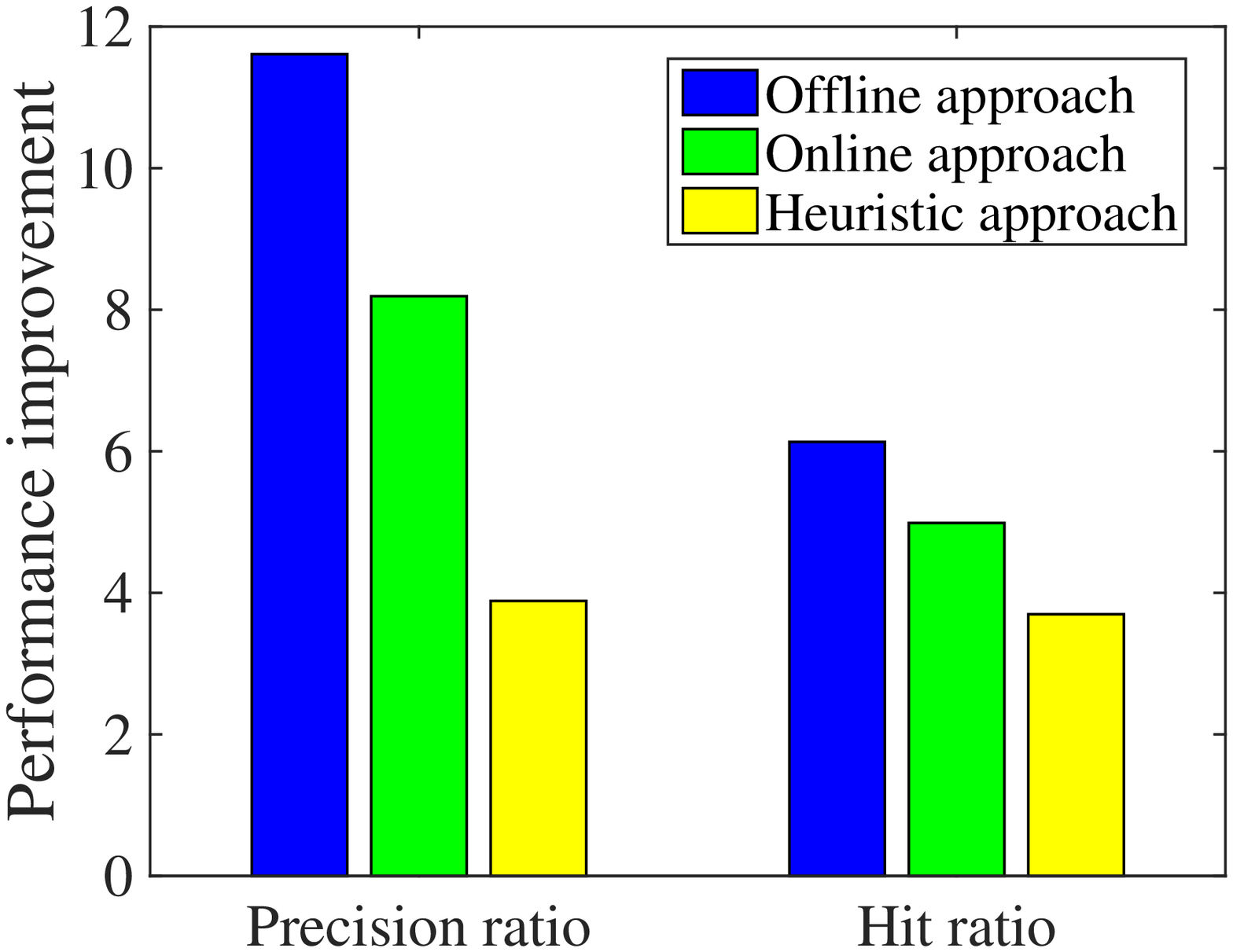}
	      \raggedright
         	  \scriptsize (b) Improvement compared with random approach.
    \end{minipage}
	\caption{Performance comparison in terms of precision and hit ratio.}
	\label{fig:prec_hit}
\end{figure}

In this part, we evaluate the effectiveness of our proposed online approach by comparing it with baselines. First, we compare these algorithms in terms of the precision ratio and the hit ratio. As shown in Fig.~\ref{fig:prec_hit}, we make the following observations: i) The online approach can achieve about $80\%$ precision accuracy, which is $9$ (resp. $1.8$) times better than the random (resp. heuristic) approach. ii) As for the hit ratio, the online approach is about $6$ (resp. $1.3$) times better than the random (resp. heuristic) approach. Note that the precision ratio of the offline algorithm is larger than $100\%$. This is reasonable because the user may repeat watching the same episodes at several time slots and a single prefetching will be regarded as several useful prefetches. iii) Both the precision ratio and the hit ratio of the online algorithm achieve better performance, about $80\%$, whereas the hit ratio of the random approach and the heuristic approach is higher than their precision ratio. This indicates that these two approaches improve the hit ratio with as many prefetchings as possible, which in turn decrease the precision ratio. iv) The performance gap between the heuristic algorithm and the random algorithm indicates the importance of users' behavior-awareness. As shown in Fig.~\ref{fig:prec_hit}(b), compared with the random algorithm, the heuristic algorithm improves both the precision accuracy and the hit ratio by $4$ times.

\begin{figure}[t]
     \begin{minipage}[t]{0.46\linewidth}
          \centering
               \includegraphics[width=\linewidth]{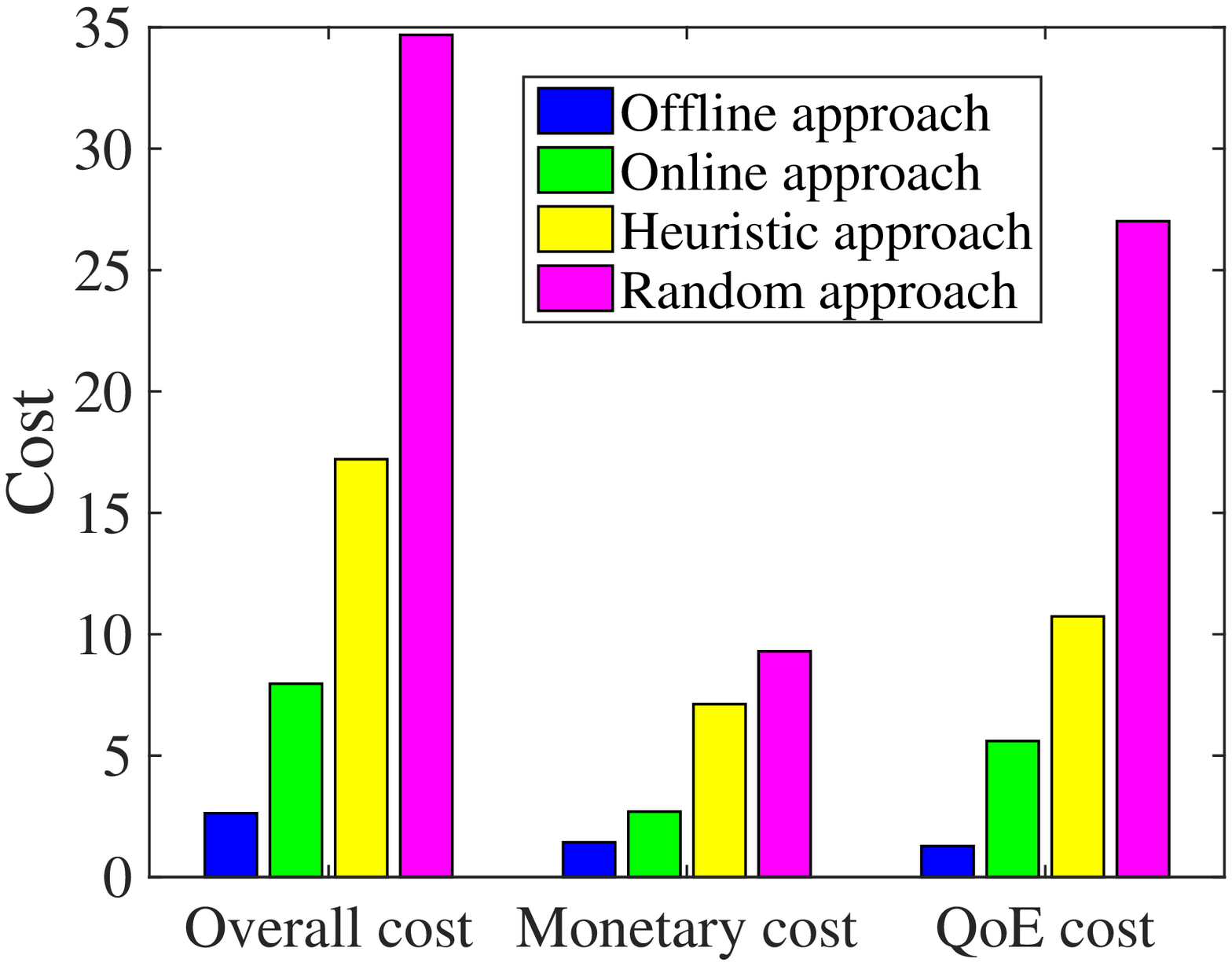}
		\raggedright
			  \scriptsize (a) Costs with different approaches.
     \end{minipage}
     \hfill
     \begin{minipage}[t]{.46\linewidth}
          \centering
			   \includegraphics[width=\linewidth]{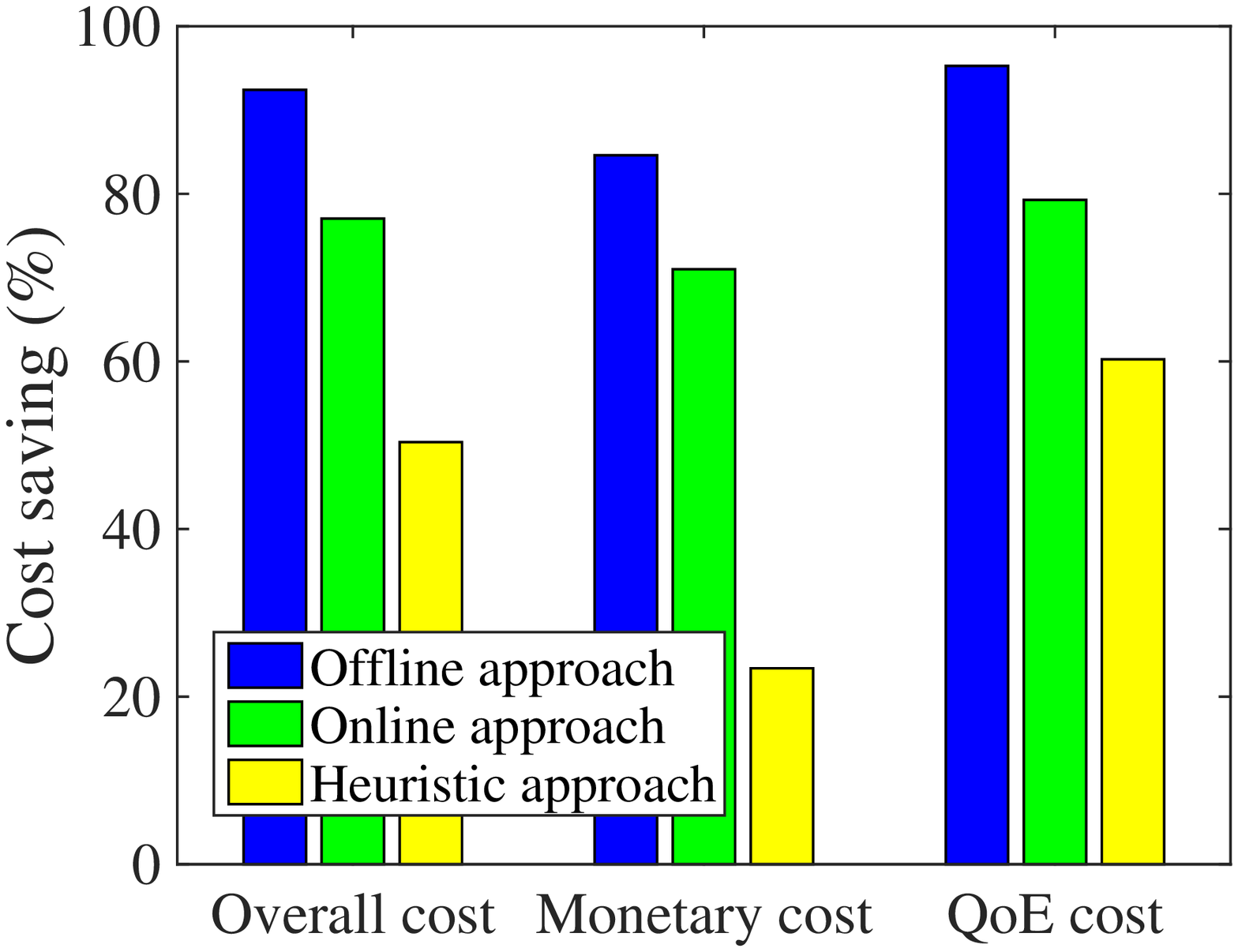}
		  \raggedright
         	  \scriptsize (b) Cost saving compared with random approach. 
    \end{minipage}
	\caption{Perfromance comparison in terms of costs.}
	\label{fig:cost}

\end{figure}

Next, we evaluate the online algorithm in terms of the different costs. We plot the overall cost, the monetary cost, as well as the QoE cost in Fig.~\ref{fig:cost}. As expected, all the costs of the online approach lies between the offline approach and the random approach. The similar monetary cost between the heuristic algorithm and random algorithm indicates the number of prefethchings of these two approaches are comparable, whereas the QoE cost gap is somewhat large, further confirming the requirement of user's watching behavior awareness. In particular, the cost saving for the online approach is slightly smaller than the offline approach. Note that there is a trade-off between the monetary cost and the QoE cost and they depend on the tuning parameter $\lambda_{1}$. Here, we analyze the performance with the fixed value $0.9$. Furthermore, the impact of $\lambda_{1}$ will be analyzed in the following section.

\subsubsection{A Case Study}
\begin{figure}[t]
  \begin{minipage}[b]{0.49\linewidth}
        \centering
               \includegraphics[width=\linewidth]{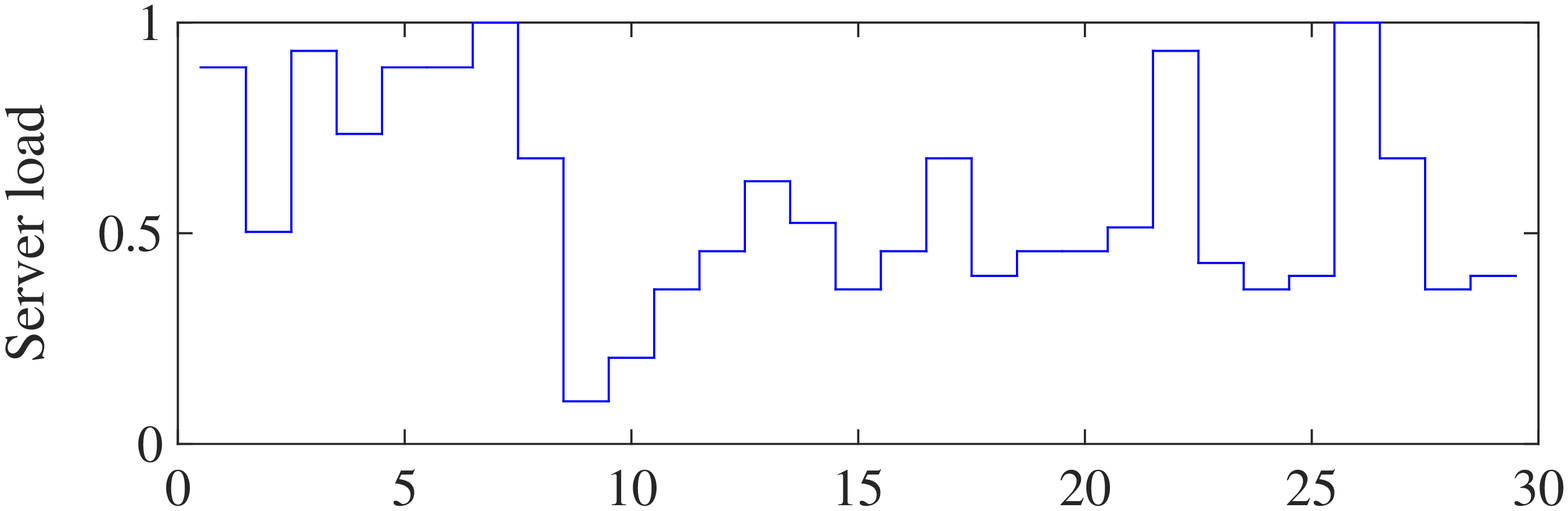}
               \raggedright
                \scriptsize (a) Server load when the user watches videos.
                \vspace{0.2cm}
  \end{minipage}%%
  \hfill
  \begin{minipage}[b]{0.49\linewidth}
    \centering
        \includegraphics[width=\linewidth]{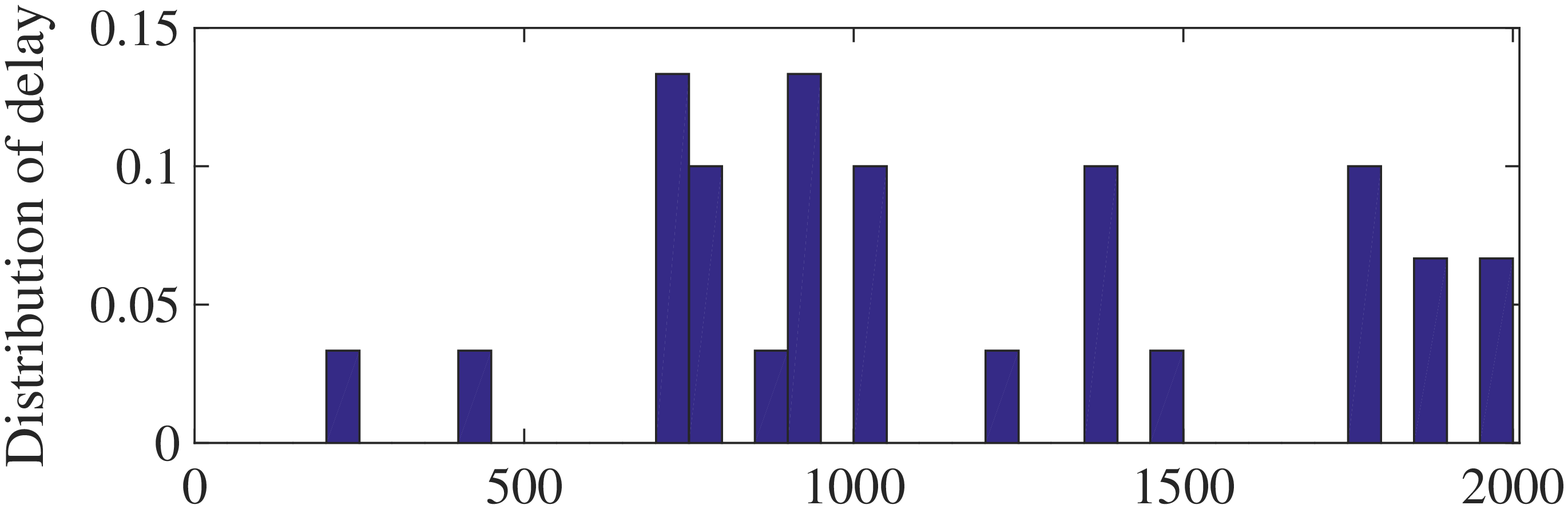}
        \raggedright
            \scriptsize (b) The startup delay without any prefetching.
            \vspace{0.2cm}
  \end{minipage}
  \vfill
  \begin{minipage}[b]{0.48\linewidth}
        \centering
               \includegraphics[width=\linewidth]{action_normal_user1.eps}
        \raggedright
            \scriptsize (c) Number of prefetched content with different strategies.
  \end{minipage}%%
  \hfill
  \begin{minipage}[b]{0.48\linewidth}
    \centering
        \includegraphics[width=\linewidth]{delay_normal_user1.eps}
        \raggedright
            \scriptsize (d) The startup delay with different prefetching strategies.
  \end{minipage}
    \caption{An example of scheduled content prefetching with different strategies in terms of the number of prefetched content and the startup delay.}
    \label{fig:normal_user}
\end{figure}

In order to dive deeper into the detailed performance, we randomly select a user case to study. We plot the number of prefetched content and the distribution of startup delay with different approaches in Fig.~\ref{fig:normal_user}. We put the server load during all periods when the user watches videos together and plot them in Fig.~\ref{fig:normal_user}(a). The histograms in Fig.~\ref{fig:normal_user}(b) is the distribution of startup delay without prefetching. Fig.~\ref{fig:normal_user}(c)(d) are the corresponding performance with different strategies. From Fig.~\ref{fig:normal_user}(c), we observe that the random approach decides to prefetch a fixed number of content (e.g.,$2$ content) regardless of the server load, whereas both the online algorithm and the offline algorithm consider the monetary cost and the QoE cost and strive to avoid prefetching at peak load periods. As for the startup delay presented in Fig.~\ref{fig:normal_user}(d), we observe that although the random approach prefetches more content (e.g., $60$ content), there are more episodes suffering from a large startup delay with the random approach than those of the online and offline approaches. Due to the space limitation, we omit the results of the heuristic algorithm.

\subsubsection{Impact of the Trade-off Parameter}
Since the parameter $\lambda_{1}$ balances the trade-off between the monetary-related cost $C^{m}$ and the QoE-related cost $C^{q}$, it is critical to design an intelligent approach which can adaptively schedule the content prefetching in response to different preferences, i.e., if the video service provider pays more attention to clients' perceived experience than the monetary cost, more content should be prefetched; otherwise, content only will be prefetched when the server is idle. We evaluate the adaptive property of our design by quantifying the monetary-related cost $C^{m}$ and the QoE-related cost $C^{q}$ with changing parameter $\lambda_{1}$. 

\begin{figure}[t]
	\centering
		\includegraphics[width=0.73\linewidth]{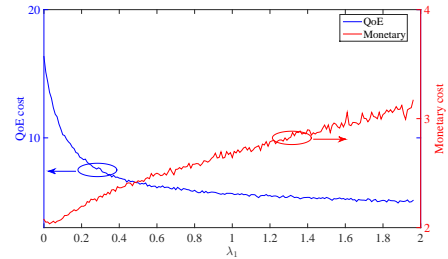}
 	\caption{The impact of trade-off parameter between the monetary-related cost and the QoE-related cost.}
 	\label{fig:lambda_impact}      
\end{figure}

As shown in Fig.~\ref{fig:lambda_impact}, we observe that with the increase of $\lambda_{1}$, the QoE cost decreases. However, as $\lambda_{1}$ increases, the monetary cost also grows, demanding a trade-off between the monetary-related and QoE-related cost in making a better content prefetching decision. The choice of $\lambda_{1}$ depends on the system budget and the desired QoE level.

\section{Related work}\label{section:relatedwork}
In this section, we survey the related work in the literature, on the edge-network content delivery and the video prefetching.

\subsection{Edge-resource Assisted Content Delivery}

Recently, some systems were developed to took advantage of the edge devices to assist the content delivery. Li et al.~\cite{li2015offline} proposed to utilize the smart APs to realize the offline downloading and claimed that the a proof-of-concept middleware could help users achieve the best expected performance by combining the advantages of cloud-based and AP-based offload downloading. Hu et al.~\cite{hu2016edge} proposed a Voronoi-like partition algorithm to take both the geo-distribution of users' request and Wi-Fi APs into account and conducted the replication in a server peak offloading manner. Ma et al.~\cite{ma2016understanding} conducted extensive measurement studies on the content placement strategies of the emerging smart router-based peer video content delivery network in China. Jayasundara et al.~\cite{jayasundara2014improving} proposed to improve the scalability of video-on-demand systems by placing the video content at the end-users' devices. Chen et al.~\cite{chen2015thunder} implemented a crowdsouring-based content distribution system, Thunder Crystal, by utilizing the resources of smart APs. They stimulated users to contribute upload bandwidth by rebating cash. 

However, these works just adopted naive content replication policies (e.g., the popularity-based or even random approach). In contrast, our work is able to make optimal content prefetching decisions in a online manner, by learning the users' watching history.

\subsection{Video Prefetching}
Prefetching video ahead of users' requests is critical to not only reduce the startup delay but also shift the traffic away during the peak periods. In this part, we classify the previous works based on the different prefetching schemes as follows.
 
\textbf{Popularity-based Video Prefetching:} Krishnappa et al.~\cite{krishnappa2011feasibility} studied the Hulu traffic in a compus network and claimed a scheme of prefetching the top-$100$ popular videos of one week was effective. Liang et al.~\cite{liang2015integrated} strived to maximize the byte-hit ratio through selecting the prefetching requests based on the number of users that were about to send the same prefetching request. 

\textbf{Social-aware Video Prefetching:} Koch et al.~\cite{koch2014optimizing} realized the video prefetching by predicting the videos a user may consume from social neighbours. Khemmarat et al.~\cite{khemmarat2012watching} and Cheng et al.~\cite{cheng2009nettube} investigated the user generated videos in YouTube and presented a prefetching scheme based on the YouTube recommendation system. Wang et al.~\cite{wang2011prefetching} proposed to reduce startup delay by predicting users' video access patterns based on both the popularity of video and the social closeness in the context of peer-to-peer video on-demand systems. Hu et al.~\cite{hu2015joint} also studied the users' video access patterns in the social community level to reduce the service latency.

\textbf{User Bahavior-aware Video Prefetching:} Grigoras et al.~\cite{grigoras2002optimizing} modeled the users' interaction with hypermedia documents based on the MDP framework and optimized video content prefetching in anticipation of user-driven navigation.  Krishnamoorthi et al.~\cite{krishnamoorthi2014quality} took advantage of parallel TCP connections to do the prefetching with a simple round-robin schedule in the context of interactive branched video streaming. They further~\cite{krishnamoorthi2015bandwidth} proposed to prefetch the alternative videos when the buffer occupancy of the video being viewed reached a threshold.
 
 Our work differentiates from them in several aspects. First, we take the dynamic time-varying server load, which is an important and practical factor, into consideration. Whereas most existing works just ignored this. Second, we systematically propose algorithms to determine the optimal number of prefetched videos, with the objective to achieve a lower accumulated cost and improved QoE. However, none of them gave such an insight. Finally, we utilize the emerging smart AP infrastructures with some prefetching algorithms to efficiently assist the content prefetching.
\section{Conclusions} \label{section:conclusion}
In this paper, we propose an AP-assisted content prefetching paradigm to balance the trade-off between the users' QoE and the incurred additional prefetching costs. Our measurement studies on users' AP connection traces and TV series session traces indicate that: i) The user's connected APs are stable over time; ii) More than $60\%$ (\emph{resp.} $90\%$) users whose over $50\%$ connections are served by their top $1$ (\emph{resp.} $5$) AP(s); iii) Users tend to watch consecutive episodes in the same TV series. Motivated by these observations, we formulate the content prefetching problem as a Markov Decision Process. Specifically, we obtain the lower and upper performance bound with a random fixed and an offline algorithm, respectively. Moreover, a heuristic algorithm also is proposed as another practical baseline.  Finally, we design a reinforcement learning algorithm, which takes both the server load and users' behavior into account, to solve the problem in the online manner. Our trace-driven experiments confirm the superiority of our learning-based approach, which not only achieves a balance between the costs and the users' QoE, but also adapts to the server load.

\bibliographystyle{IEEEtran}
\bibliography{acm_mylib_short}

\end{document}